\documentclass[aps,prl,11pt,onecolumn]{revtex4}
\usepackage{amsmath}
\usepackage{amssymb}
\usepackage{bbm}
\usepackage{graphicx}
\usepackage{framed}
\usepackage{color}
\usepackage{hyperref}
\usepackage{epstopdf}
\usepackage{dsfont}
\usepackage{amssymb}
\usepackage{amsmath}
\usepackage{amsthm}
\usepackage{wrapfig}
\usepackage{relsize}
\usepackage{bm}
\usepackage[latin1]{inputenc}
\usepackage{enumitem}
\usepackage{multirow}
\usepackage{pifont}

\newcommand{\Ket}[2]{\vert#1\rangle_{#2}}
\newcommand{\ket}[1]{\vert#1\rangle}

\newcommand{\ave}[1]{\langle#1\rangle}

\begin{document}

\title{Interaction-free, single-pixel quantum imaging with undetected photons}

\author{Yiquan Yang$^{1}$, Hong Liang$^{1}$, Xiaze Xu$^{1}$, Lijian Zhang$^{1}$, Shining Zhu$^{1}$, and Xiao-song Ma$^{1,2,3}$\footnote{Xiaosong.Ma@nju.edu.cn}}

\affiliation{1. National Laboratory of Solid State Microstructures, School of Physics and Collaborative Innovation Center of Advanced Microstructures, Nanjing University, Nanjing 210093, China \\
2. Synergetic Innovation Center of Quantum Information and Quantum Physics, University of Science and Technology of China, Hefei 230026, China \\
3. Hefei National Laboratory, Hefei 230088, China}

\begin{abstract}
A typical imaging scenario requires three basic ingredients: 1. a light source that emits light, which in turn interacts and scatters off the object of interest; 2. detection of the light being scattered from the object and 3. a detector with spatial resolution. These indispensable ingredients in typical imaging scenarios may limit their applicability in the imaging of biological or other sensitive specimens due to unavailable photon-starved detection capabilities and inevitable damage induced by interaction. Here, we propose and experimentally realize a quantum imaging protocol that alleviates all three requirements. By embedding a single-photon Michelson interferometer into a nonlinear interferometer based on induced coherence and harnessing single-pixel imaging technique, we demonstrate interaction-free, single-pixel quantum imaging of a structured object with undetected photons. Thereby, we push the capability of quantum imaging to the extreme point in which no interaction is required between object and photons and the detection requirement is greatly reduced. Our work paves the path for applications in characterizing delicate samples with single-pixel imaging at silicon-detectable wavelengths.
\end{abstract}

\maketitle

\section{Introduction}
Over the past few decades, several imaging protocols based on quantum technologies have been realized\cite{genovese2016real,moreau2019imaging}, which have expanded the application capabilities of optical imaging. These include ghost imaging (GI)\cite{pittman1995optical,strekalov1995observation}, quantum imaging with undetected photons (QIUP)\cite{lemos2014quantum}, and interaction-free measurements (IFMs)\cite{dicke1981interaction,elitzur1993quantum}. The quantum GI scheme relies on the spatial correlations of entangled photon pairs and requires two-photon coincident measurements. Furthermore, ghost imaging can also be realized with classical intensity-fluctuation correlations\cite{shapiro2012physics}. Later, various single-pixel imaging (SPI) protocols were proposed\cite{shapiro2008computational,duarte2008single,altmann2018quantum,edgar2019principles,gibson2020single}, where the spatial correlations are not between two photons but between one photon and a programmable mask held in a spatial light modulator (SLM).

In contrast to modern digital cameras employing array sensors to capture images, SPI use a sequence of masks to interrogate the scene along with the correlated intensity measurements by a single-pixel detector. The spatially resolved masks are usually generated by computer and displayed by SLM. Combined with compressive techniques\cite{duarte2008single}, the number of sampling measurements is fewer than the total number of pixels in the image. Thereby, SPI can reduce the data processing requirement, and shows potential capability for high dimensional sensing\cite{edgar2019principles}. On the other hand, the modern single-photon detector is featured by improved detection efficiency, lower dark counts, and faster timing response\cite{hadfield2009single}. Such enhancements have significance to applying SPI into weak signal detection scenarios, such as scattering medium imaging or long-range 3D imaging\cite{altmann2018quantum}.

The QIUP scheme is based on induced coherence (IC), which was first proposed by Zou, Wang, and Mandel\cite{zou1991induced}. They used two photon sources to generate photon pairs. By overlapping path of two sources for one photon (idler)\cite{zou1991induced,wang1991induced,herzog1994frustrated} and establishing the so-called path identity\cite{krenn2017entanglement,hochrainer2021quantum},  there is no information about the origin of the other photon (signal). Thus, the signal photon is in the superposition state of being created in either of the sources. The phase and transmissivity of the idler photon are encoded in the interference of the signal photon. Inserting one object onto the idler path between two sources, one can obtain images exclusively with the signal photons which have no interaction with the object\cite{lemos2014quantum}. In contrast to GI, QIUP does not involve the detection of the photon illuminating the object or any coincidence measurement. This is a advantage of QIUP, as the wavelength of the detected photon can be chosen independently from that of the photon interacting with the object\cite{lemos2014quantum}. This concept was further explored in infrared (IR) spectroscopy\cite{kalashnikov2016infrared}, optical coherence tomography\cite{valles2018optical,paterova2018tunable}, mid-IR imaging\cite{kviatkovsky2020microscopy,paterova2020hyperspectral,paterova2020quantum}, terahertz (THz) sensing\cite{kutas2020terahertz}, biological microscopy\cite{buzas2020biological}, and holography\cite{topfer2022quantum}. Recently, the related SU(1,1) interferometer has been investigated and employed in quantum-enhanced metrology\cite{yurke19862,hudelist2014quantum,chekhova2016nonlinear,ou2020quantum,du20222}.

However, both GI and QIUP require direct interaction between the object and the probe photon. By contrast, IFM allows one to detect the presence of a photon-supersensitive object without direct interaction. The concept of IFM was first proposed by Dicke\cite{dicke1981interaction}. Elitzur and Vaidman (EV) extended this idea and proposed harnessing the wave-particle duality of single photons to realize IFM\cite{elitzur1993quantum}. In the EV scheme, the Mach-Zehnder interferometer (MZI) is aligned to have destructive interference at the dark output in the absence of the object. As one opaque object was placed onto either path of the interferometer, the presence of the object modifies the optical interferograms of the MZI. Any photon detection event at the dark port indicates the photon comes from the path not containing the object. Hence, the measurements for binary objects were deemed interaction-free. For grey or quantum objects, Kwiat et. al. considered `quantum interrogation' as a more appropriate terminology\cite{kwiat1999high}. High-efficiency IFMs have been realized by using the discrete quantum Zeno effect\cite{kwiat1999high,kwiat1995interaction,ma2014chip}. Later, exploiting the advantages of the lithographically written waveguides, highly efficient IFMs with an integrated chip have been realized\cite{ma2014chip}. Furthermore, the concept of IFM was also applied to quantum imaging\cite{white1998interaction,zhang2019interaction,hance2021counterfactual} and quantum communication\cite{noh2009counterfactual,liu2012experimental,salih2013protocol,cao2017direct,alonso2019trace,salih2022laws}.

\begin{figure}
\centering
\includegraphics[width=0.5\textwidth]{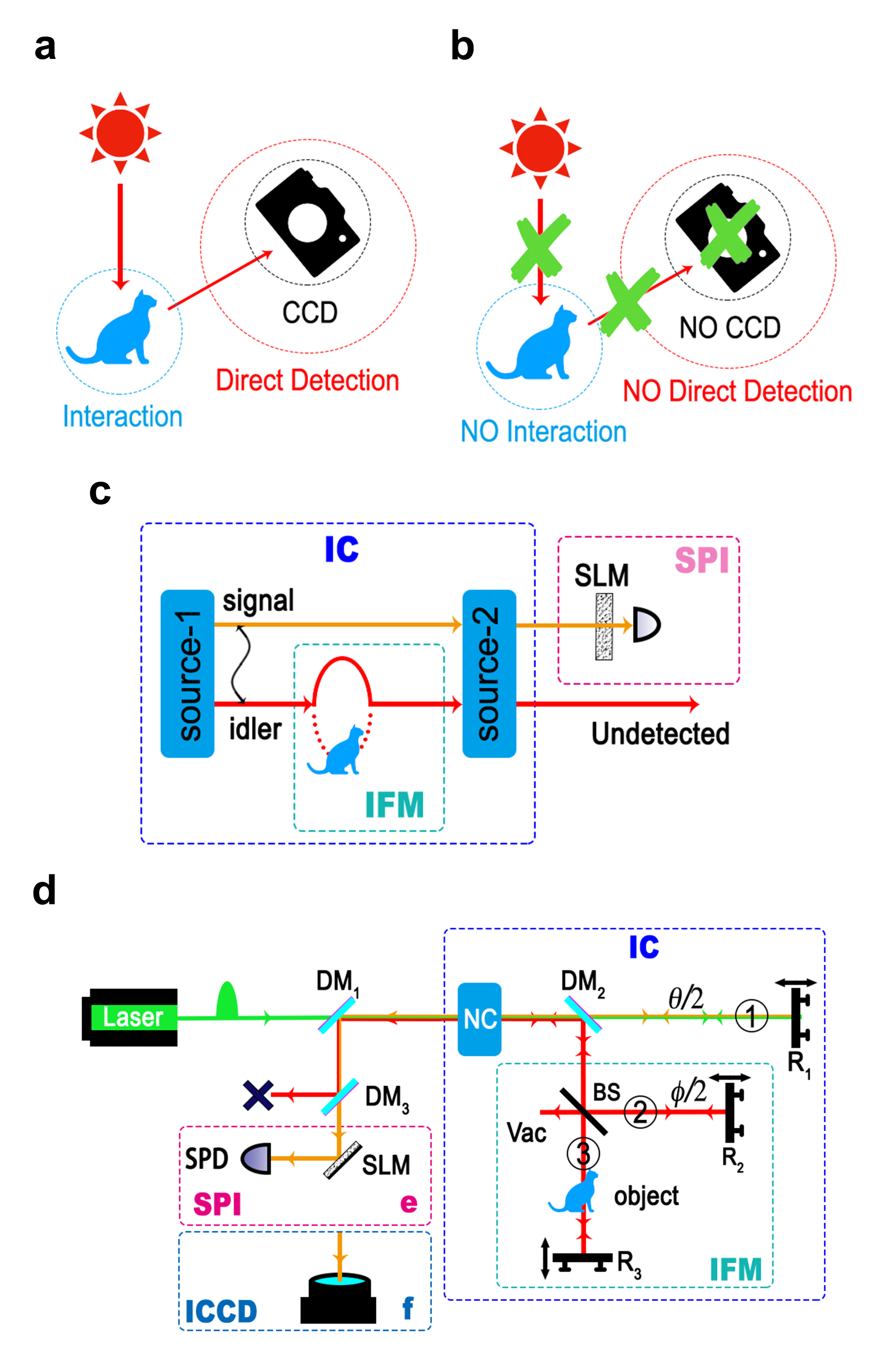}
\caption{Conceptual scheme and setup of interaction-free, single-pixel quantum imaging with undetected photons. \textbf{a} Classical imaging. Classical imaging requires a light source that emits light, which in turn interacts with the object of interest; a detector with spatial resolution directly detects light scattered from the object. \textbf{b} In our imaging scheme, all the above conditions are alleviated, which means imaging can be realized without direct interaction, direct detection, and a charge-coupled device (CCD) camera. \textbf{c} Conceptual diagram. Our imaging protocol is realized by integrating two interferometers based on the induced coherence (IC) and the interaction-free measurement (IFM) with the single-pixel imaging (SPI). \textbf{d} Experimental setup. A folded version of (\textbf{c}) realizes IC. The double-pass spontaneous parametric down-conversion processes correspond to source-1 and source-2 in panel \textbf{c}. The IFM is realized with a single-photon Michelson interferometer. The object is placed into arm \ding{174} of the IFM. Vac denotes the vacuum port of the IFM module. The idler photon is filtered out by dichroic mirror DM$_3$ and remains undetected throughout the entire imaging process. The circled numbers represent different arms of the interferometer. $\theta$ and $\phi$ represent the phase of the signal photon and the relative phase of the IFM. We also perform interaction-free quantum imaging with an intensified CCD (ICCD) camera (shown in panel \textbf{f}), instead of using SPI (shown in panel \textbf{e}). IC: induced coherence. IFM: interaction-free measurement; SPI: single-pixel imaging; SLM: spatial light modulator; DM: dichroic mirror; NC: nonlinear crystal; BS: beam splitter; R: reflector; SPD: single-photon detector.}
\label{concept}
\end{figure}

Above-mentioned quantum imaging protocols can alleviate one or two specific requirements inherent in typical imaging. The goal of our work is to develop and experimentally demonstrate a quantum imaging protocol, in which all essential requirements of a typical imaging process (Fig. \ref{concept}a) are simultaneously alleviated: 1. physical interaction --- photons emitted from a light source impinge on the object and interact with it; 2. direct detection --- scattered photons are detected to reveal the presence of object; 3. spatially resolved detector, such as a charge-coupled device (CCD) --- spatial information is acquired. Here we report the realization of interaction-free, single-pixel quantum imaging with undetected photons (Fig. \ref{concept}b) by combining IFM\cite{dicke1981interaction,elitzur1993quantum}, SPI\cite{shapiro2008computational,duarte2008single,altmann2018quantum,edgar2019principles,gibson2020single}, and the IC\cite{zou1991induced,wang1991induced,herzog1994frustrated,krenn2017entanglement,hochrainer2021quantum}interferometer, as shown in Fig. \ref{concept}c. Based on the principle of QIUP, the probe (idler) photon remains undetected throughout the imaging process. Information about the object of interest (in Fig. \ref{concept}c, a cat) carried by the idler photon is transferred to the signal photon. However, unlike the previous QIUP realizations\cite{lemos2014quantum,kviatkovsky2020microscopy,paterova2020hyperspectral,paterova2020quantum}, IFM removes the requirement that the probe photon and the object interact. Finally, spatial information is obtained by SPI instead of using a CCD. SPI consists of the SLM and the single-pixel detector.

We have pushed the current quantum imaging ability to the extreme point where the detection requirements are minimized and both the illuminating photon and detected photon have no interaction with the object. Thus, our imaging protocol has the potential to characterize fragile or photon-sensitive systems, such as biological tissues\cite{stephens2003light} and quantum states of atomic ensembles\cite{wolfgramm2013entanglement,eckert2008quantum}. Furthermore, spatially resolved single-photon detection is inefficient, costly and even unavailable in the challenging wavelength ranges, such as those in far-IR or even longer wavelength ranges. We can overcome the above limitations via minimum detection requirements. By employing economic, low-noise and high-efficiency visible single-photon detectors, we can realize long-wavelength interaction-free imaging.

\section{Results}
\subsection{The experimental scheme}
The experimental setup is shown in Fig. \ref{concept}d. It consists of three main parts, the IC interferometer, the IFM interferometer, and the SPI module. The IC interferometer is formed by double-passing a nonlinear crystal (NC) in a folded Michelson geometry\cite{chekhova2016nonlinear}. As the pump passes the NC twice, it can generate a single pair of signal and idler photons via a forward (from left to right) or backward (from right to left after being reflected by the mirror R$_1$) spontaneous parametric down-conversion (SPDC) process. The forward and backward SPDC processes correspond to source-1 and source-2 in Fig. \ref{concept}c, respectively. The pump laser with a wavelength of $\lambda_p=532\,nm$ (green) collinearly generates from the NC a pair of frequency-nondegenerate correlated photons with wavelengths of $\lambda_s=810\,nm$ (yellow) and $\lambda_i=1550\,nm$ (red, the subscripts $s$ and $i$ denote the signal and idler photon). Both photon-pair sources are based on the type-I SPDC process, i.e., a horizontally polarized pump photon converts into vertically polarized signal and idler photons. We use a lithium niobate crystal (with thickness $L = 2\,mm$ and cutting angle $\theta=68^{\circ}$ for type-I phase matching) as the NC. Various dichroic mirrors (DMs) are used to separate and combine frequency-nondegenerate photons. DM$_1$ transmits the pump and reflects both signal and idler photons. DM$_2$ transmits both pump and signal photons and reflects idler photons. DM$_3$ transmits signal photons and reflects idler photons. By adjusting the relative optical delays between the signal photon, the idler photon, and the pump with three motorized translation stages mounted on the mirrors R$_1$, R$_2$, and R$_3$, we establish path identity\cite{krenn2017entanglement,hochrainer2021quantum} of both forward and backward SPDC processes (see the Supplementary Note 1). Thereby, we obtain the quantum interferometer enabled by IC.

As shown in Fig. \ref{concept}d, the idler photon generated from the forward SPDC process is reflected by DM$_2$ and injected into the IFM module, realized by a Michelson interferometer. If the object is not in the IFM module, then this idler photon will deterministically exit from the Vac port, providing the relative phase $\phi$ of the IFM module to be 0. Consequently, no forward-generated idler photon goes back to the NC, and hence IC interference is prohibited. In this case, no IC interference of the signal and idler photons can be observed. On the other hand, if an opaque object is inserted into arm \ding{174} of the IFM module, then the mere presence of this object will prohibit the interference of the IFM. With the probability of 25\%, the idler photon generated in the forward SPDC will propagate back into the IC interferometer. Note that this 25\% portion accounts for the idler photon passing the balanced beam splitter (BS) twice along arm \ding{173} of the IFM module. Thus, the path identity of the IC interferometer can be partially established, and hence signal photons show interference with 50\% visibility. We can then infer the presence of the object by evaluating the interferogram of signal photons. We emphasize that the probe (idler) photon has never interacted with the object throughout the entire process. This is a subtlety of IFMs, at whose basis lies the wave-particle duality of a single photon. A single photon is indivisible because of its particle property and cannot split on a beam splitter\cite{grangier1986experimental}. Therefore, every idler photon that goes back to the IC interferometer must not have propagated through arm \ding{174} of the IFM and hence would not have interacted with the object. 

The single-pixel imaging with the signal photon is realized in the SPI module (see Fig. \ref{concept}d). In SPI, one can reconstruct the multi-pixel image by interrogating the image with a set of spatially resolved masks and simultaneously recording the correlated intensity with a single-pixel detector. Mathematically, the image can be described by $\boldsymbol{I}=\boldsymbol{P}\cdot\boldsymbol{T}$, where $\boldsymbol{I}$ represents the pixelated image, $\boldsymbol{P}$ is a set of masks, and $\boldsymbol{T}$ is a collection of correlated photon counts (weighting) for the corresponding mask set. A commonly used set of spatially resolved masks is the Hadamard set\cite{pratt1969hadamard,sloane1976masks}. We use a SLM to display the Hadamard mask to project the image onto a single-mode fibre-coupled single-photon detector (SPD). The single-pixel SPD records the signal-photon counts for each mask.

\begin{figure}
\includegraphics[width=0.8\textwidth]{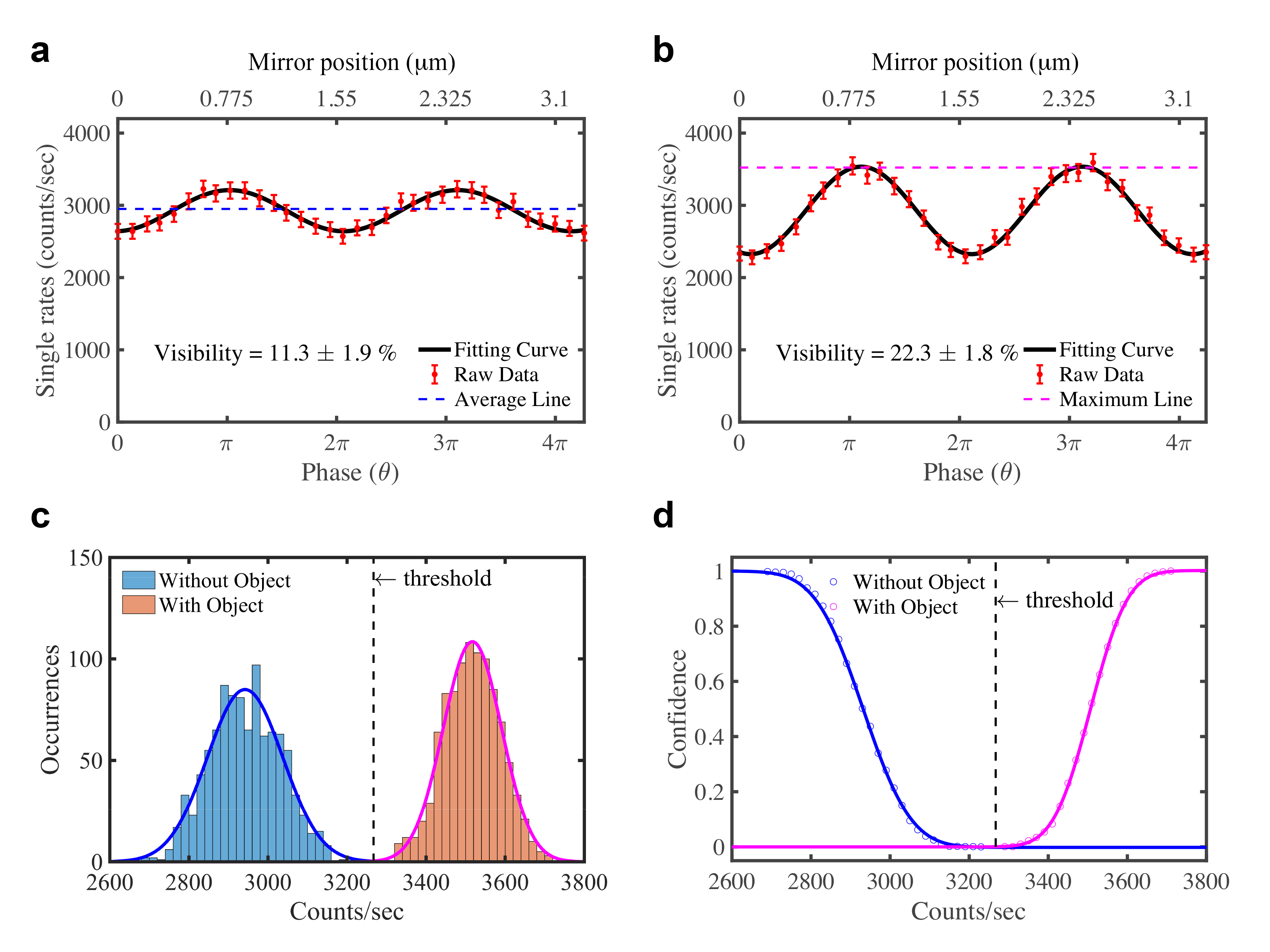}
\caption{Interaction-free quantum sensing with undetected photons. By adjusting the phase $\theta$ of the signal photons, we obtain (\textbf{a}) the residual interference of the signal photon in the absence of the object with a visibility of $11.3\pm1.9$\%. \textbf{b} When the object is present, the idler photon performs a successful IFM and induces the interference of the signal photon. The experimental visibility is $22.3\pm1.8$\%. The relative phase $\theta$ can be tuned by adjusting optical path difference of two arms in the nonlinear interferometer with motorized translation stages mounted on the mirrors R$_1$. \textbf{c} Counting histogram of the signal photon fitted by two Gaussian distributions. The width of each bin is 20 counts per second. The mean counts per second for sensing the presence (purple) and absence (blue) of the object are about 3,500 and 2,950, respectively. By setting a threshold (vertical black dash line), we clearly distinguish whether the object is present or not. \textbf{d} The confidence of the sensing is obtained by integrating the two Gaussian distributions shown in (\textbf{c}); a confidence level above 99.93\% at the 3.4-sigma threshold is obtained. Error bars in panels \textbf{a} and \textbf{b} indicate two standard errors of the mean.}
\label{Sensing}
\end{figure}

\subsection{Quantum Sensing}

As the first step of our experiment, we realized interaction-free quantum sensing with undetected photons by inserting an opaque object into the IFM module. In the absence of the object, the visibility of the IFM module was $93.9\pm1.1$\% (100\% for the ideal case). This imperfection of the IFM interference allowed the idler photon to go back to the IC interferometer with a low probability of about 2\% at the destructive phase point $\phi=0$. Imperfections of the IFM and other optical components in our experimental setup resulted in residual interference with small visibility for the signal photon, which was about $11.3\pm1.9$\% (Fig. \ref{Sensing}a). In the ideal case, this visibility of signal interference should be zero. Once the opaque object was inserted into arm \ding{174} of the IFM module, a fraction of the idler photons generated by the forward SPDC were reflected back to the NC with no interaction with the object (see explanations above). These idler photons induced interference of signal photons with a visibility of $22.3\pm1.8$\% (Fig. \ref{Sensing}b). Therefore, the signal photon had different visibilities, depending on the presence of the object. Operationally, we use the fitted maximum counts ($\theta=\pi$) of the IC interference (purple dashed line in Fig. \ref{Sensing}b) to denote a successful detection of the object with IFM and use the averaged counts (blue dashed line in Fig. \ref{Sensing}a) to denote the absence of the object. The single counts of the signal photon thus allow us to decide whether or not the object is in the IFM module without interacting with it. See the Supplementary Note 1 for details of the interference visibilities.

In Fig. \ref{Sensing}c, we show the counting histogram of the signal photon for interaction-free quantum sensing without detecting the probe photon (idler), fitted with two well-separated Gaussian functions. We clearly distinguish the presence (counts above threshold) and absence (counts below threshold) of the object with a confidence level above 99.93\% at the 3.4-$\sigma$ threshold (indicated with the vertical black dash line in Fig. \ref{Sensing}d).

\subsection{Quantum imaging}

\begin{figure}
\includegraphics[width=0.8\textwidth]{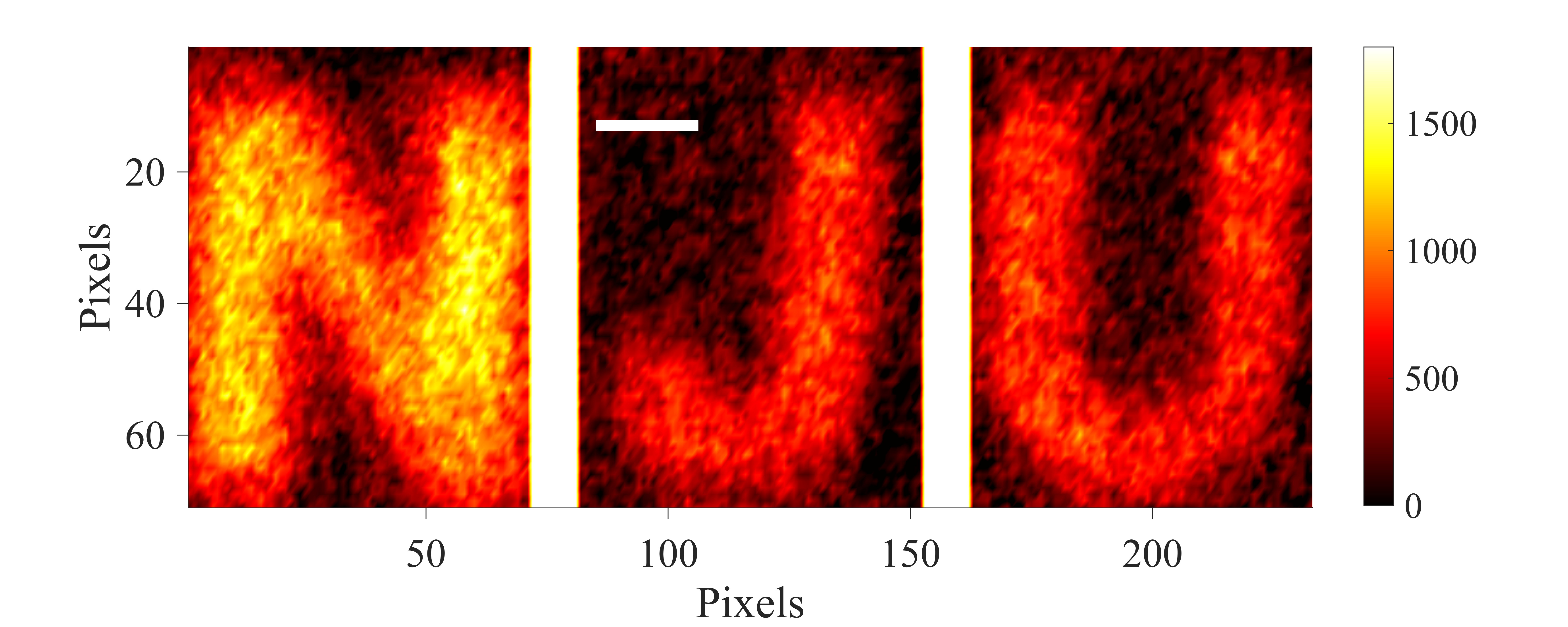}
\caption{Interaction-free quantum imaging result of the `\textbf{NJU}' characters with undetected photons. In the presence of the plate in arm \ding{174} of the interaction-free measurement module, two images are obtained with ICCD at phase setting ($\theta=\pi$, $\phi=0$) and ($\theta=0$, $\phi=\pi$) (see the SI). The difference between these two images (see the SI) is shown above. The signal photon inside the three characters has higher visibility than in outside regions. The three characters `\textbf{N}', `\textbf{J}' and `\textbf{U}' are measured independently. Length of the scale bar (white colour) in the figure is 200 $\mu m$.}
\label{NJU_ICCD}
\end{figure}

Next, we realized interaction-free imaging with undetected photons (IFIUP). We employed an intensified CCD camera (ICCD, Andor-iStar, DH334T-18U-73, shown in Fig. \ref{concept}f) to image a spatially structured object, which is a 3D-printed `\textbf{NJU}' logo (the abbreviation of \textbf{N}an\textbf{j}ing \textbf{U}niversity). The region with the characters is transparent and the complementary region is opaque. Therefore, the character and the remaining regions correspond to the absence/presence of the object, respectively. These two regions have different interference visibilities. Using the method presented above, we can obtain different counts for these two regions and perform IFIUP, which is recorded with the ICCD. At the same phase setting as in previous quantum sensing ($\theta=\pi$, $\phi=0$, see Fig. \ref{concept}d), we obtained one image of the `\textbf{NJU}' plate. Then, we adjusted the relative phase $\phi$ to $\pi$, and the signal phase $\theta$ to 0. At this phase setting, the idler photon deterministically propagates back into NC. The signal photon correlated with the idler photon propagating along the character region interferes constructively at this phase setting ($\theta=0$, $\phi=\pi$). In Fig. \ref{NJU_ICCD}, we show the subtraction of the obtained images for these two settings, highlighting the difference between constructive and destructive interference and enhancing the interference contrast. The size of each pixel in the image is $13\,\mu m \times 13\,\mu m$. For the details of spatial resolution and bandwidth, please refer to the SI. For the details of imaging resolution, spatial properties of photon pairs and other discussions, please refer to the Supplementary Note 2-6.

\begin{figure}
\includegraphics[width=0.8\textwidth]{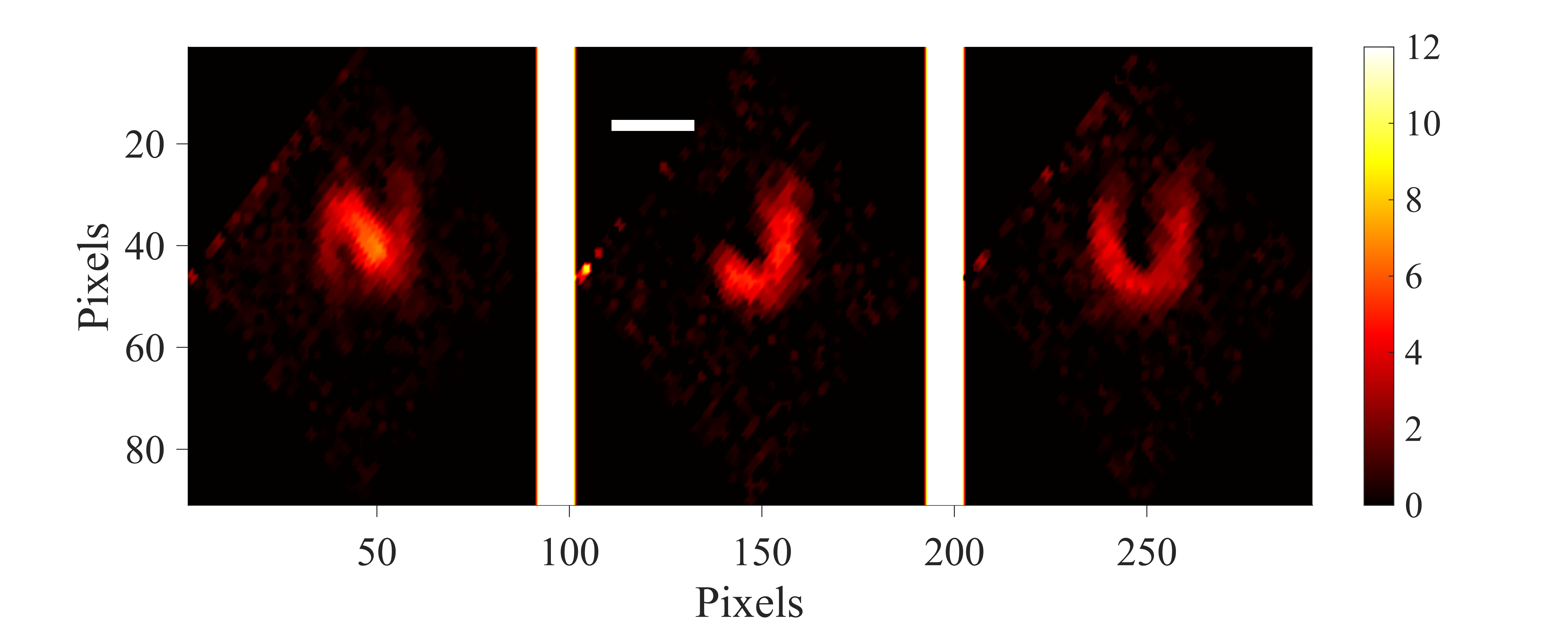}
\caption{Interaction-free, single-pixel quantum imaging result of the `\textbf{NJU}' plate with undetected photons. The three characters `\textbf{N}', `\textbf{J}', and `\textbf{U}' are measured independently with the single-pixel imaging technique. Note that in our experiment, the micromirror of the SLM is rotated $45^{\circ}$ at a mount. The edge of SLM can be vaguely seen at the upper left corner of the image. We rotate the reconstructed images by $45^{\circ}$ to obtain the above images. The background noise in the signal-photon counts results in some speckles in the reconstructed images. Because of the structural difference of characters and limited detection area of the fiber-coupled detector, the visual result of character `\textbf{N}' is less prominent to `\textbf{J}' and `\textbf{U}'. Length of the scale bar (white colour) in the figure is 500 $\mu m$.}
\label{NJU_SPI}
\end{figure}

Finally, using the SPI module shown in Fig. \ref{concept}(e), we realized the interaction-free, single-pixel quantum imaging with undetected photons. With the SPI method, the object can be reconstructed by multiplying each mask in the sampling set by the corresponding signal-photon counts, resulting in a set of weighted masks that can be summed up to form an image. Here, we employed a SLM device (ViALUX, V-650LNIR) to display the Hadamard masks\cite{pratt1969hadamard,sloane1976masks}, with a pixel number of $64\times 64$ (the size of each pixel is $32.4 \,\mu m\times32.4\, \mu m$). We employed 1024 masks to perform SPI. For each mask, we recorded signal-photon counts at four settings, $C(\theta=\pi, \phi=0)$, $C(\theta=0, \phi=0)$, $C(\theta=0, \phi=\pi)$, and $C(\theta=\pi, \phi=\pi)$ and determined for each mask the signal-photon count $C_M$ ((see the Supplementary Note 4 and 7)):
\begin{equation}
C_M = C(\theta=0, \phi=\pi) - C(\theta=\pi, \phi=\pi) - C(\theta=\pi, \phi=0) + C(\theta=0, \phi=0).
\end{equation}
Each mask is multiplied by the corresponding value of $C_M$ to give a set of weighted masks. By summing up all weighted masks, we reconstruct the images of the `\textbf{NJU}' plate, shown in Fig. \ref{NJU_SPI}. 

\section{Discussion}

We have demonstrated interaction-free, single-pixel quantum imaging with undetected photons, which is fundamentally different with typical imaging. By harnessing the wave-particle duality of single photons, interaction-free imaging is performed with the probe (idler) photons. Based on induced coherence, the image information carried by the idler photons, which have no interaction with the object, is transferred to the signal photons. Finally, using the SPI method, we reconstruct the image of the object solely with a sequence of signal-photon counts detected by a visible single-photon detector without spatial resolution. Therefore, our imaging protocol has alleviated all three requirements in the typical imaging scenario. We have pushed the capability of quantum imaging to the extreme point.  Both the illuminating photon and probe photon have no physical interaction with the interested object. Furthermore, the detection requirement in quantum imaging is minimized to visible single-photon detectors without spatial resolution.

We note that the IFM efficiency can be further improved by using more advanced optical interferometric setups and low-loss optical switches\cite{kwiat1995interaction,kwiat1999high,ma2014chip}. In addition, broadband phase-matching conditions in nonlinear materials can be exploited to generate highly frequency-nondegenerate photon pairs, providing flexibility in the spectral range for both signal photon and idler photon. Thereby, our protocol can be extended into IR\cite{kviatkovsky2020microscopy,paterova2020hyperspectral,paterova2020quantum} or THz\cite{kutas2020terahertz} single-pixel imaging without disturbing the sample using low-noise and high-efficiency visible single-photon detectors. 
We emphasize that the above statement of `interaction-free' is only sensible for binary objects\cite{white1998interaction,zhang2019interaction,hance2021counterfactual} due to the interference property of IFM. For grey or quantum objects, our experimental facility has the potential to realize single-pixel quantum interrogation\cite{kwiat1999high} with undetected photons. Compared with conventional imaging, both interaction-free imaging and quantum interrogation based on single-photon interference can reduce sample exposure to intense light fields. Our measurement protocol may be beneficial for the probe of sensitive specimens, which is limited by unavailable photon-starved detection capabilities\cite{kviatkovsky2020microscopy} and inevitable damage induced by interaction\cite{putnam2009noninvasive,turner2021interaction,taylor2016quantum}. 
As such, we hope this work will stimulate wide research interest in multiple fields, such as delicate material investigation\cite{wolfgramm2013entanglement,eckert2008quantum} and life science\cite{taylor2016quantum}.

\section{Acknowledgments}

We thank Anton Zeilinger for the inspiring discussion. This research was supported by the National Key Research and Development Program of China (2017YFA0303704, 2019YFA0308704), the National Natural Science Foundation of China (Grant No. 11690032), the NSFC-BRICS (No. 61961146001), the Leading-Edge Technology Program of Jiangsu Natural Science Foundation (No. BK20192001), the Fundamental Research Funds for the Central Universities, and the Innovation Program for Quantum Science and Technology (Grant No. 2021ZD0301500).

\bibliography{reference}

\clearpage
\newpage

%

\widetext

\section*{Supplementary information: Interaction-free, single-pixel quantum imaging with undetected photons}

%
%

\subsection*{Supplementary Note 1. Interference analysis}
Our imaging setup (Fig. 1(d) in the main text) integrates a nonlinear interferometer\cite{Chekhova} based on induced coherence and an interaction-free measurement (IFM) module. In this folded geometry, a pump laser illuminates a nonlinear crystal (NC) twice in sequence. Photon pairs generated in the first and second pass through the NC are denoted by $\Ket{1}{s_1}\Ket{1}{i_1}$ and $\Ket{1}{s_2}\Ket{1}{i_2}$, respectively. By adjusting the position of mirrors R$_1$, R$_2$, and R$_3$, we establish the path identity\cite{Krenn} ($\Ket{1}{s_1}\Ket{1}{i_1}\to \Ket{1}{s_2}\Ket{1}{i_2} = \Ket{1}{s}\Ket{1}{i} $). In the absence of object in the interferometer, the obtained state after the second pass is
\begin{equation}\label{no loss state}
\ket{\Psi} = \alpha \ket{0} + \beta \bigg\{ \Ket{1}{s} \Ket{1}{i} + e^{i\theta}\Ket{1}{s} \big[ (T-e^{i\phi}R)\Ket{1}{i} + i\sqrt{TR}(1+e^{i\phi})\Ket{1}{i_0} \big] \bigg\}.
\end{equation}
In our experiment, NC is a low-gain nonlinear crystal weakly pumped by laser. In this case, the probability of down-conversion is so low ($\beta \ll 1, \alpha \approx 1$) that only one pair of photons is generated via a forward or backward spontaneous parametric down-conversion (SPDC) process. $\Ket{1}{i_0}$ denotes the idler photon emitted from the vacuum (Vac) port. T and R are the transmissivity and reflectivity of the idler photon on the beam splitter (BS), respectively.

\begin{figure}
\includegraphics[width=1\textwidth]{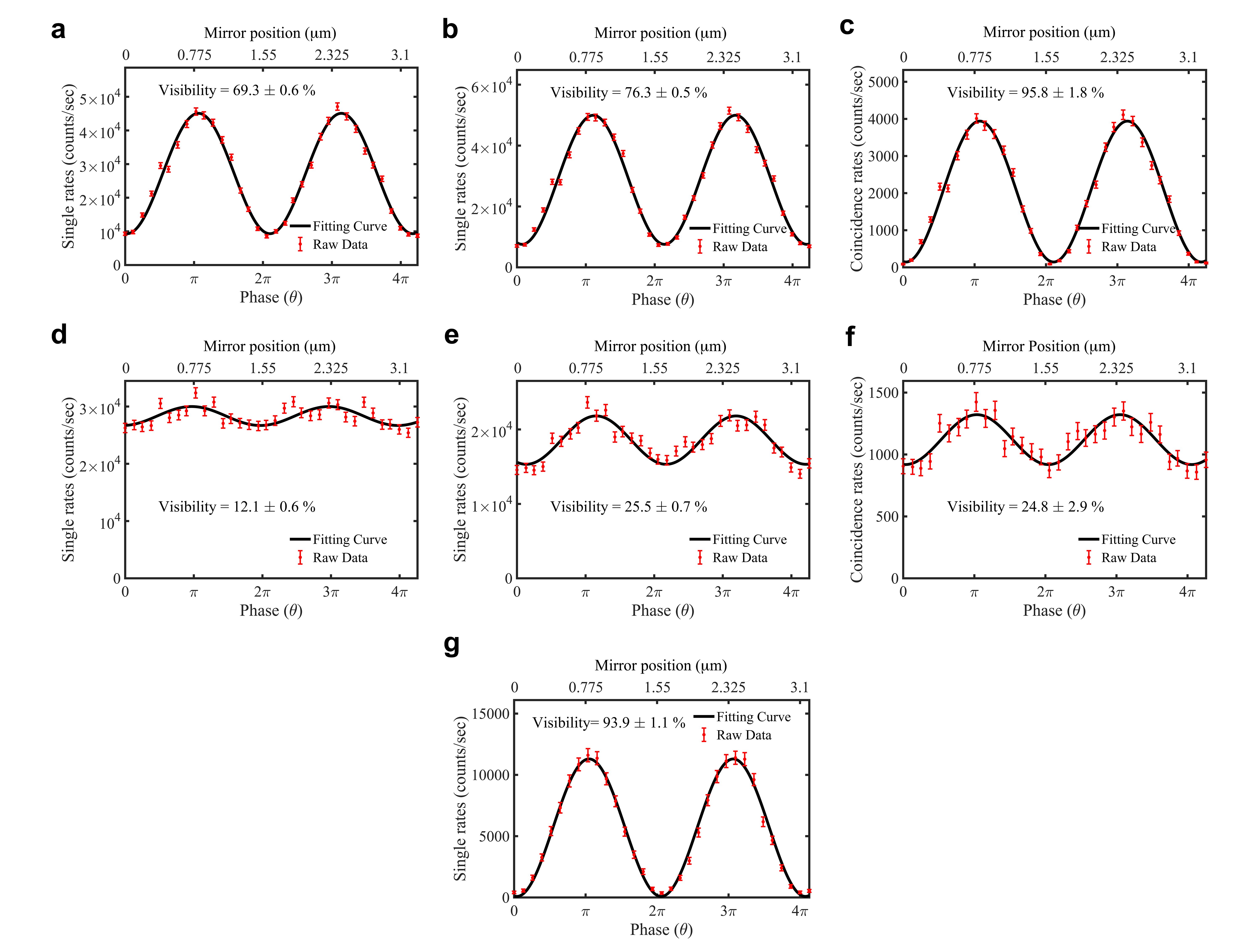}
\caption{Measurement of interference patterns. The observed interference patterns of the signal photon, idler photon, and coincidence counts at the constructive phase point $\phi=\pi$ (\textbf{a}-\textbf{c}) and the destructive phase point $\phi=0$ (\textbf{d}-\textbf{f}) of the interaction-free measurement (IFM). \textbf{g} The observed interference pattern of the IFM module. Error bars in panels \textbf{a}, \textbf{b}, \textbf{d}, and \textbf{e} indicate five standard errors of the mean, in panels \textbf{c} and \textbf{f} indicate two standard errors of the mean.}
\label{interference_pattern}
\end{figure}

Based on Eq. (\ref{no loss state}), we obtain the theoretical visibilities of the signal photon, idler photon, coincidence counts and IFM compared with the experimental values, as shown in Table \ref{visibility}. The measured interference patterns of the signal photon, idler photon, coincidence counts and IFM are shown in Fig. \ref{interference_pattern}.

\begin{table}[htbp]
\begin{tabular}{|c|c|c|c|c|c|c|c|}
\hline
\multirow{2}{*}{} & \multicolumn{2}{c|}{signal photon}                   & \multicolumn{2}{c|}{idler photon}                    & \multicolumn{2}{l|}{coincidence counts}              & \multirow{2}{*}{IFM}    \\ \cline{2-7} 
                  & \multicolumn{1}{c|}{at $\phi=\pi$}      & \multicolumn{1}{c|}{at $\phi=0$} & \multicolumn{1}{c|}{at $\phi=\pi$}      & \multicolumn{1}{c|}{at $\phi=0$} & \multicolumn{1}{c|}{at $\phi=\pi$}      & \multicolumn{1}{c|}{at $\phi=0$} &        \\ \hline
theoretical results     & 100\%  & 0                      & 100\%  & 0                      & 100\%  & 0                      & 100\%  \\ \hline
experimental results & $\approx$ 69.3\% & $\approx$ 12.1\%                 & $\approx$ 76.3\% & $\approx$ 25.5\%   &              $\approx$ 95.7\% & $\approx$ 24.8\%                 & $\approx$ 93.9\% \\ \hline
\end{tabular}
\caption{Comparison of visibilities between the theoretical and experimental results. The visibilities of the signal photon, idler photon, coincidence counts and interaction-free measurement (IFM) module in the theoretical and experimental case. $\phi$ is the relative phase of the IFM.}
\label{visibility}
\end{table}

If the relative phase $\phi$ of the IFM is $\pi$, the idler photon generated in the forward SPDC process deterministically propagates back into the NC. Therefore, the interference visibilities of the signal photon, idler photon, coincidence counts are 100\%. Assuming $T=R=1/2$, the visibility of lossless IFM also reaches 100\%. However, at the phase point $\phi=0$, the idler photon no longer propagates back into the NC. Therefore, interference is prohibited. Here, because we do not consider optical propagation losses and the imperfection of path alignment in our experimental setup, the theoretical visibilities take higher values than observed ones (Table \ref{visibility}). The visibility of the signal photon is close to the value reported by Lemos et al. in their quantum-imaging experiment\cite{Lemos}.

\subsection*{Supplementary Note 2. Imaging resolution analysis}

We consider the edge-spread function (ESF) to study the sharpness of our imaging system\cite{Harms,Fuenzalida}. The ESF is given by $I(x) = a - b\times \text{erf}[(x-x_c)/\sigma]$, where the erf is the error function and the $\sigma$ is the width of the edge and defined as
\begin{equation} \label{sigma}
\sigma = \frac{f_s\lambda_s}{\sqrt{2}\pi\omega_p} M.
\end{equation}

\begin{figure}
\includegraphics[width=0.6\textwidth]{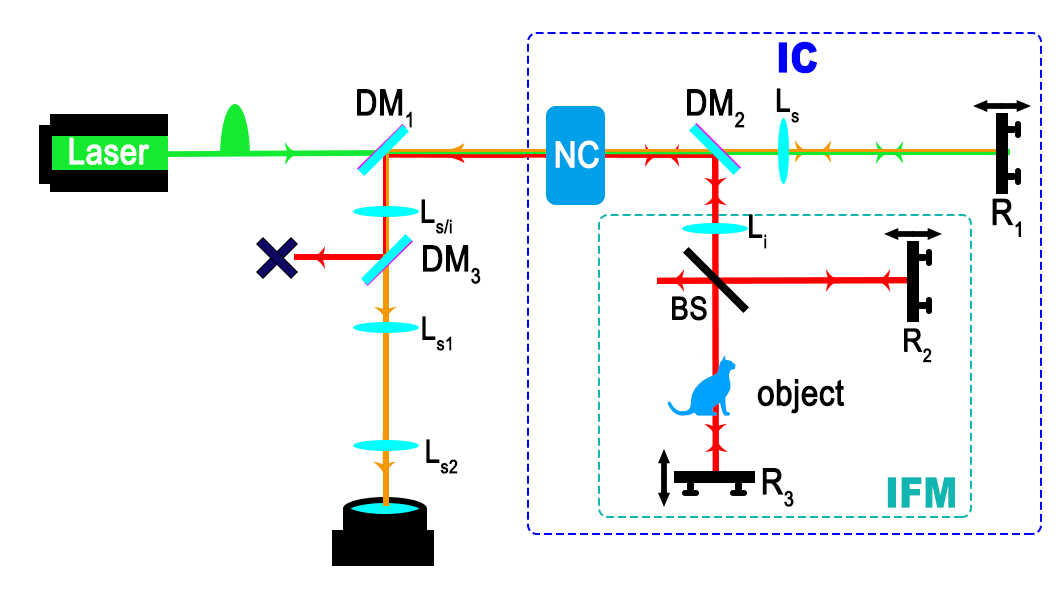}
\caption{Description of all imaging lenses in the experimental setup. The confocal lenses L$_s$ and L$_i$ with focal length $f$ = 100 mm in the signal and idler arms project the momentum spectra of down-converted photons onto the reflective mirrors (R$_1$, R$_2$ and R$_3$) and the object. The nonlinear crystal (NC) and the object are placed at a focal distance from the lens L$_i$. Furthermore, both L$_s$ and L$_i$ make spatial modes of forward and backward generated photon pairs indistinguishable at the nonlinear crystal. L$_s$ ensures the same pump waist size of the forward and backward SPDC processes at the nonlinear crystal. The confocal three-lens system L$_{s/i}$, L$_{s1}$ and L$_{s2}$ map the transverse momentum spectra of the signal onto the camera (image) plane. The focal lengths of three-lens system are $f_{s/i}$ = 100 mm, $f_{s1}$ = 250 mm and $f_{s2}$ = 100 mm, respectively. The 4f lens system (L$_{s1}$-L$_{s2}$) is used to form the reimaging image from the lens L$_{s/i}$ and determines magnification ratio of the final signal image. For other details of the experimental setup, please refer to Fig. 1d in the main text.}
\label{lens}
\end{figure}

\begin{figure}
\includegraphics[width=0.7\textwidth]{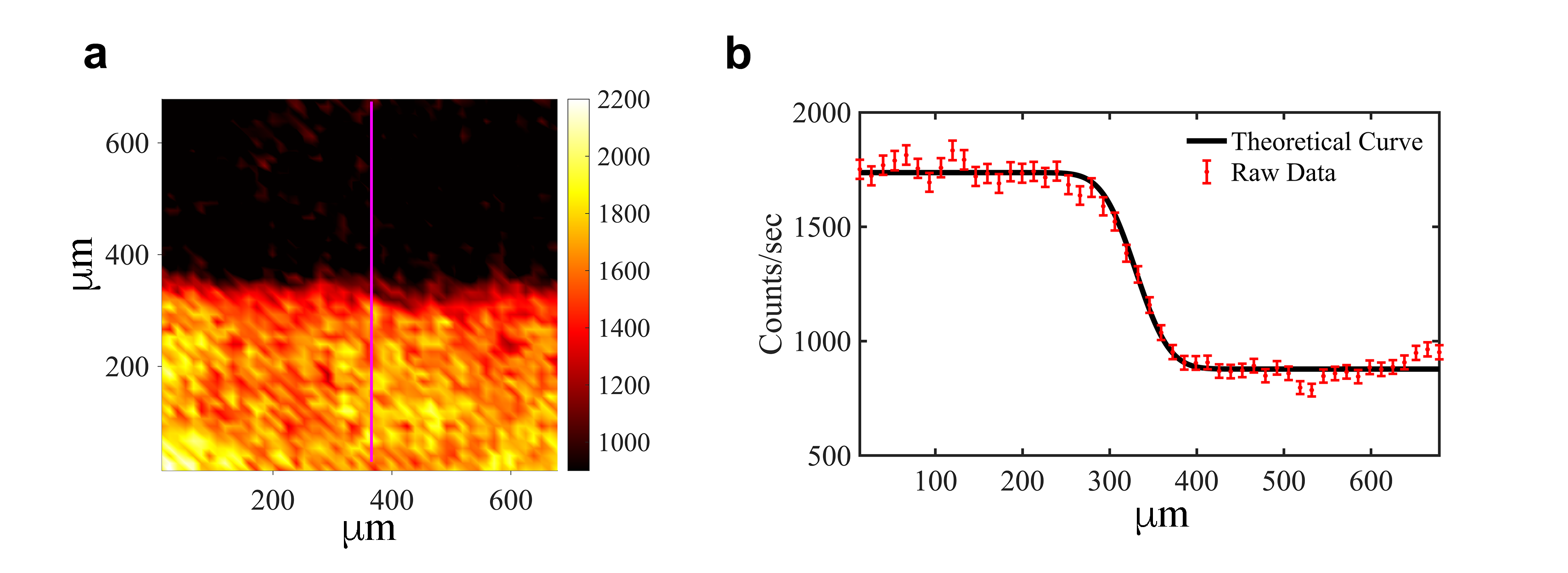}
\caption{Measurement of imaging resolution. \textbf{a} The image of the sharp knife-edge measured by the intensified CCD camera. \textbf{b} The signal-photon counts along the purple line in panel \textbf{a}. The black line is the curve of the theoretical edge-spread function with $\sigma = 43 \mu m$.}
\label{resolution}
\end{figure}

$M$ is the magnification ratio of the 4f imaging system before the intensified charge-coupled device (ICCD) camera, which is about 0.4. The focal length of the lens on the arm of the signal photon ($\lambda_s \approx 810 \,nm$) is $f_s \approx 100 \,mm$. The detailed description of all imaging lenses in our experimental setup are illustrated in Fig. \ref{lens}. The pump waist on the NC is $\omega_p \approx171 \,\mu m$. Entering the above parameters into Eq. \ref{sigma}, we derive that the theoretical value of $\sigma$ is about $43 \,\mu m$. Experimentally, the width of the edge, $\sigma$, can be obtained by imaging a sharp knife-edge. The result is shown in Fig. \ref{resolution}a. The experimentally measured ESF data (red) is shown in Fig. \ref{resolution}b and agrees well with the theoretical prediction (black). The resolution of the idler photon can be further promoted by increasing the pump waist or decreasing the focal length of the lens.

\subsection*{Supplementary Note 3. Spatial properties of photon pairs}

In this section, we discuss spatial properties of the correlated photon pairs generated by the SPDC process. Considering the pump is one Gaussian laser beam:
\begin{equation} 
E(\bm{k}_p) \propto \exp \bigg(-\frac{ |\bm{k}_{p}|^2 \omega_p^2}{4}\bigg),
\end{equation}
where $\bm{k}_p$ is the transverse momentum of the pump, $\omega_{p}$ is beam waist size. The mode function of photon pair\cite{Gilaberte} is described by
\begin{equation} 
\Phi(\bm{k}_s,\bm{k}_i) \propto \exp \bigg(-\frac{ |\bm{k}_{s}+\bm{k}_{i}|^2 \omega_p^2}{4}\bigg) \text{sinc} \bigg(-\frac{\Delta\bm{k}L}{2}\bigg),
\end{equation}
which is determined by the interplay of the spatial properties of the pump beam and the phase matching function. $\bm{k}_{s}$ and $\bm{k}_{i}$ represent transverse momentum of signal and idler. $\Delta\bm{k}$ is the momentum mismatch of the SPDC process, $L$ denotes the crystal length. The momentum correlation of signal-idler photon pairs is determined by the pump waist, which is approximately 171 $\mu m$.

To determine the imaging capacity of our implementation\cite{Erkmen,Kviatkovsky}, we also need to calculate the anticipated field of view (FoV), which is given by:
\begin{equation}
\sigma_{i} = \frac{f_i \lambda_i}{\sqrt{2} \pi \omega_p},
\end{equation}
where $\lambda_i$ is the wavelength of the idler. The number of spatial modes (spatial bandwidth) is therefore:
\begin{equation}
N_i = \bigg( \frac{\text{FoV}}{\sigma_i} \bigg)^2 \approx 948.
\end{equation}
Because the SPDC source in our experiment is spatially broadband, we do not need raster scanning to image objects.

\subsection*{Supplementary Note 4. Imaging data acquisition and processing}

\begin{figure}
\includegraphics[width=0.4\textwidth]{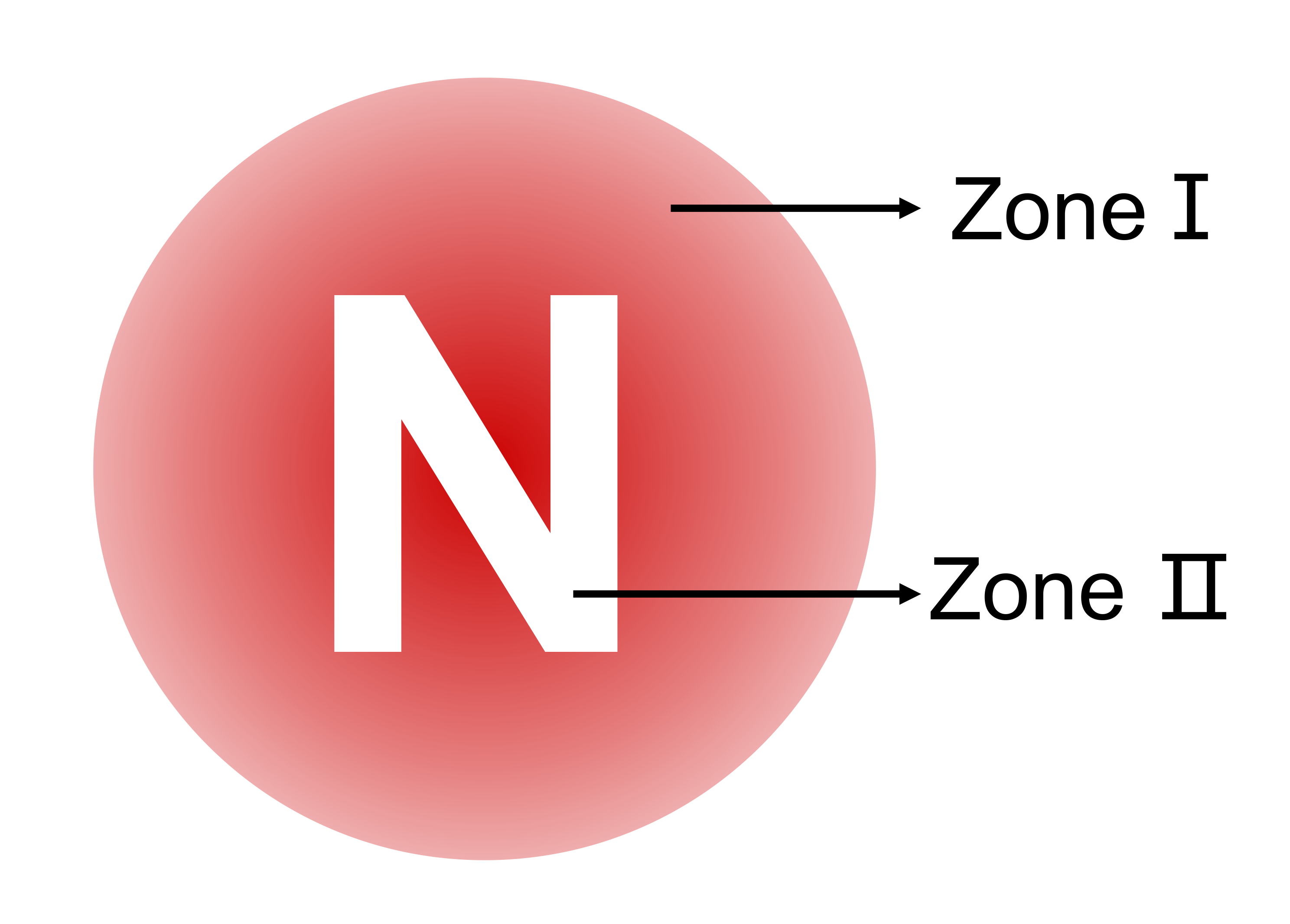}
\caption{The Imaging plate. It can be divided into opaque Zone I and transparent Zone II. Here we only use the character `\textbf{N}' to illustrate the imaging plate. Other characters follow the same rules.}.
\label{character}
\end{figure}

For the `\textbf{NJU}' plate used in our imaging experiment, the character region is transparent (Zone II), and the complementary region (Zone I) is opaque, as shown in Fig. \ref{character}. If this plate is inserted into arm \ding{174} of the IFM, the signal photons in the two regions will have different interference visibilities. Based on Eq. (\ref{no loss state}), the interference image of Zone I and Zone II can be described as
\begin{gather}
C_{I}(\theta,\phi)= P_{I(x,y)}\bigg[ 1-\frac{1}{2}\cos(\theta+\phi) \bigg],\\
C_{II}(\theta,\phi)= P_{II(x,y)}\bigg\{ 1+ \frac{1}{2} \big[ \cos\theta - \cos(\theta+\phi) \big]\bigg\}.
\end{gather}
where $P_{I(x,y)}$  and $P_{II(x,y)}$ are zone-dependent photon-emission rates of the NC in Zone I and Zone II, respectively.

At the constructive interference point of the IFM, $\phi=\pi$, the entire interference image of the signal photon can be represented as
\begin{equation}
\begin{split}
C(\theta,\phi=\pi) &= C_I(\theta,\phi=\pi)+C_{II}(\theta,\phi=\pi) \\
&= P_{I(x,y)}\bigg[ 1+\frac{1}{2}\cos\theta \bigg] + P_{II(x,y)}\big[ 1+ \cos\theta \big].
\end{split}
\end{equation}

By tuning the phase $\theta$, we can obtain the constructive and destructive images:
\begin{gather}
C_{max}(\theta,\phi=\pi)= C(\theta=0,\phi=\pi)=\frac{3}{2}P_{I(x,y)} + 2P_{II(x,y)}; \label{cons_max}\\\
C_{min}(\theta,\phi=\pi)= C(\theta=\pi,\phi=\pi)=\frac{1}{2}P_{I(x,y)}.
\end{gather}
At the destructive interference point of the IFM, $\phi=0$, the entire interference image of the signal photon can be expressed as
\begin{equation}
\begin{split}
C(\theta,\phi=0) &= C_I(\theta,\phi=0)+C_{II}(\theta,\phi=0) \\
&= P_{I(x,y)}\bigg[ 1 - \frac{1}{2}\cos\theta \bigg] + P_{II(x,y)}.
\end{split}
\end{equation}
In this case, the constructive and destructive images are
\begin{gather}
C_{max}(\theta,\phi=0)= C(\theta=\pi,\phi=0)=\frac{3}{2}P_{I(x,y)} + P_{II(x,y)}; \label{des_max}\\
C_{min}(\theta,\phi=0)= C(\theta=0,\phi=0)=\frac{1}{2}P_{I(x,y)} + P_{II(x,y)}.
\end{gather}
To enhance the interference contrast of the imaging, we subtract Eq. (\ref{des_max}) from Eq. (\ref{cons_max}) to obtain the final image:
\begin{equation}\label{data_process_ICCD}
C_{max}(\theta,\phi=\pi) - C_{max}(\theta,\phi=0) = P_{II(x,y)}.
\end{equation}
Then only the signal counts in the character region are left on the image. This data-processing method is employed in interaction-free quantum imaging with undetected photons by ICCD (shown in Fig. 3 in the main text). In this method, we need to record the images of the signal photon at two settings, ($\theta=0, \phi=\pi$) and ($\theta=\pi, \phi=0$). 

For the interaction-free, single-pixel quantum imaging with undetected photons, we employ the following method to process the imaging data: 
\begin{equation}\label{data_process_SPI}
C_{max}(\theta,\phi=\pi)-C_{min}(\theta,\phi=\pi)-C_{max}(\theta,\phi=0)+C_{min}(\theta,\phi=0) = 2P_{II(x,y)}.
\end{equation}
In this case, we need to record signal-photon counts for each Hadamard mask at four settings, ($\theta=0, \phi=\pi$), ($\theta=\pi, \phi=\pi$), ($\theta=\pi, \phi=0$) and ($\theta=0, \phi=0$). This method can enhance the brightness of the image because the signal-photon counts obtained in Eq. (\ref{data_process_SPI}) is twice as much as Eq. (\ref{data_process_ICCD}). Processing the four recorded signal-photon counts based on Eq. (\ref{data_process_SPI}), we get the corresponding counts of each Hadamard mask. Each mask in the sampling set is then multiplied by the corresponding counts to give a set of weighted masks that can be summed up to reconstruct the image\cite{Gibson}.

\subsection*{Supplementary Note 5. Interaction-free quantum imaging with undetected photons}
\begin{figure}
\includegraphics[width=0.6\textwidth]{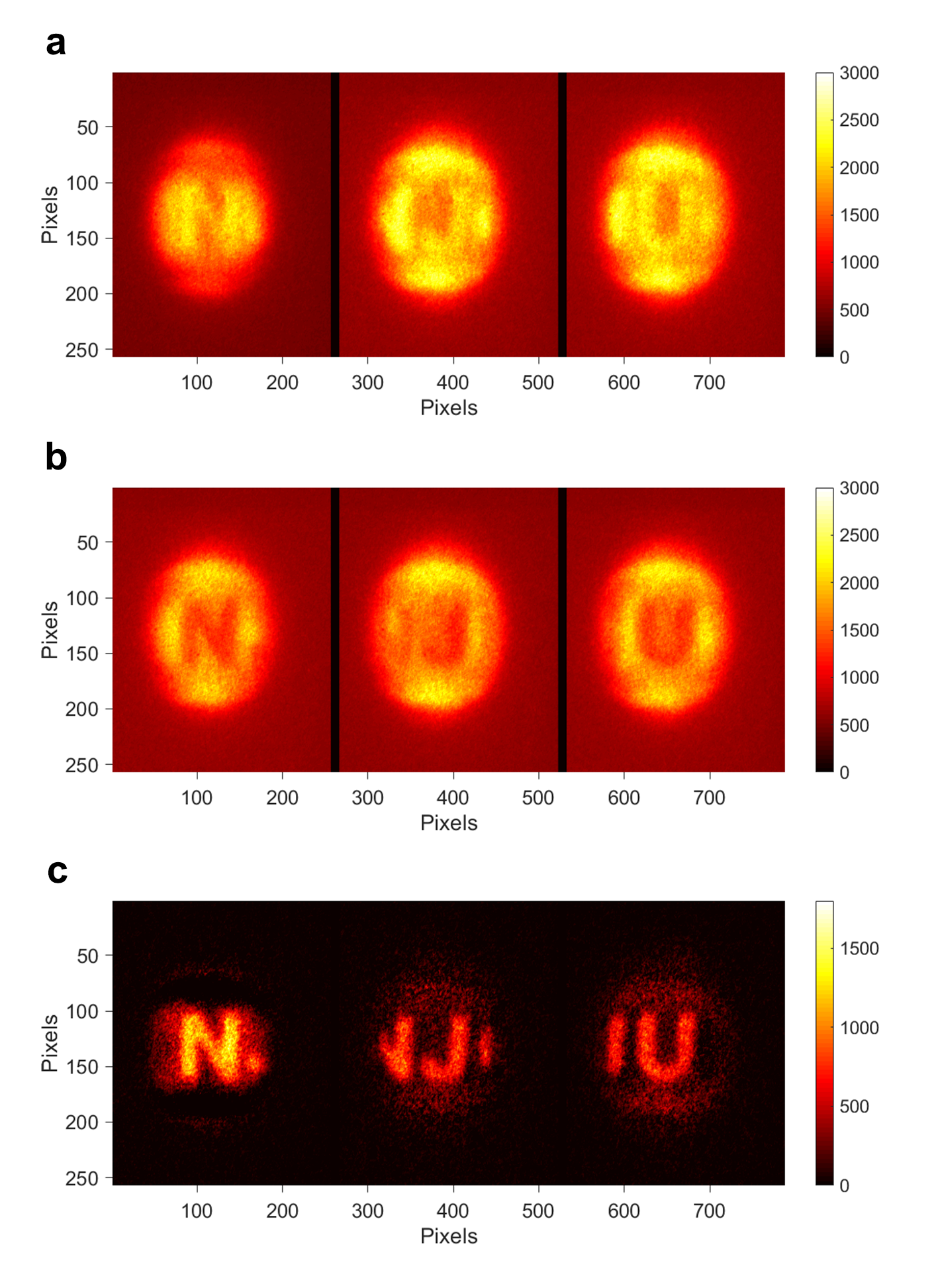}
\caption{Interaction-free quantum imaging results with undetected photons by intensified CCD. \textbf{a} and \textbf{b} are obtained at ($\theta=0, \phi=\pi$) and ($\theta=\pi, \phi=0$) settings, respectively. (c) is the subtraction of two images in panel \textbf{a} and \textbf{b}, which leads to an enhancement of the interference contrast.}
\label{NJU_ICCD_raw}
\end{figure}

\begin{figure}
\includegraphics[width=0.5\textwidth]{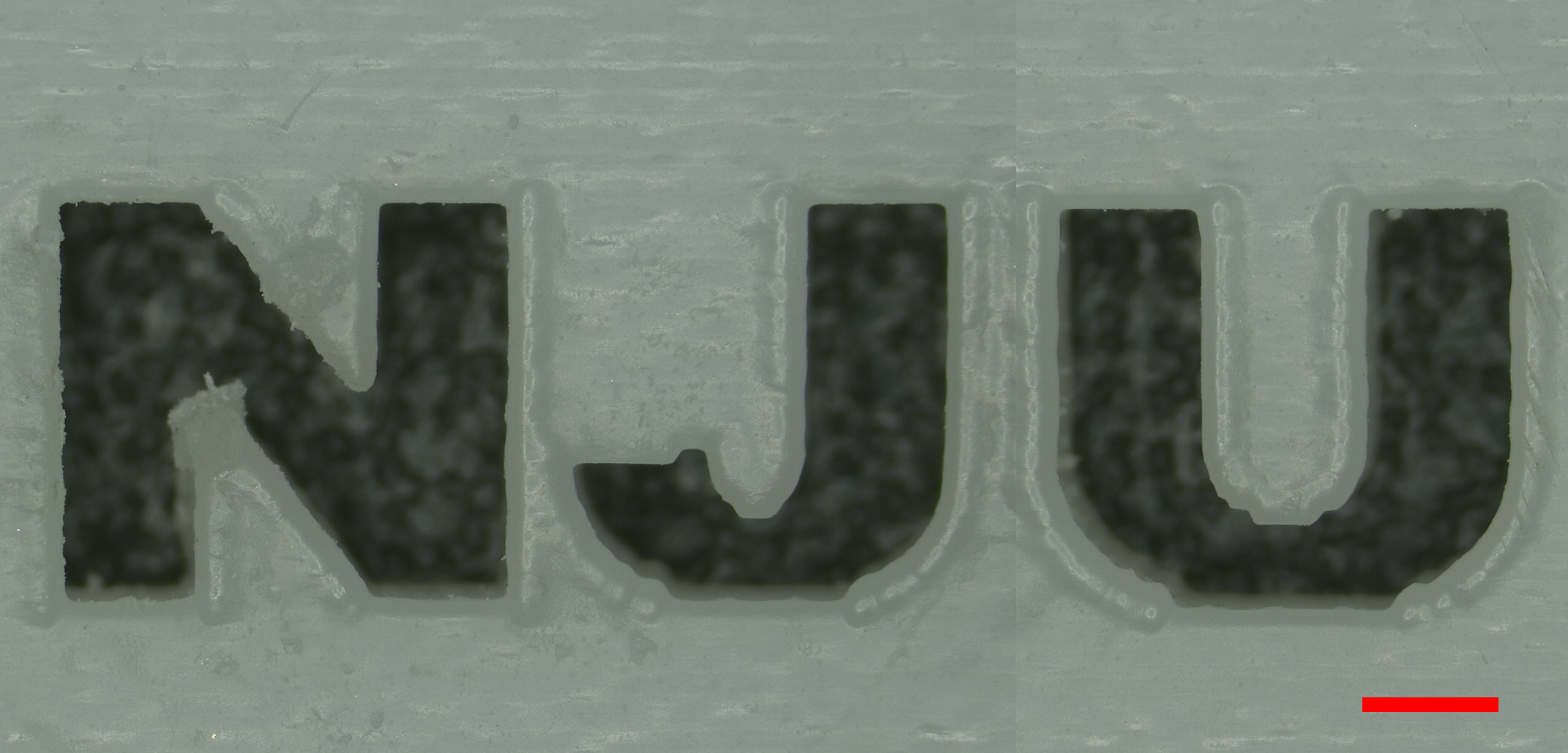}
\caption{Ground truth of the `\textbf{NJU}' plate taken using an optical microscope. Length of the scale bar (red colour) in the figure is 1 mm.}
\label{NJU_microscope}
\end{figure}

Using the ICCD, we recorded the image of the `\textbf{NJU}' plate at these two settings, ($\theta=0, \phi=\pi$) and ($\theta=\pi, \phi=0$), as shown in Fig. \ref{NJU_ICCD_raw}a and \ref{NJU_ICCD_raw}b. All data shown here are raw data. The difference between both images is shown in Fig. \ref{NJU_ICCD_raw}c. Due to the defocused lens, we observe the circular fringes outside the character region\cite{Hochrainer}. The ground truth of the `\textbf{NJU}' plate is shown in Fig. \ref{NJU_microscope}.

\subsection*{Supplementary Note 6. Single-pixel quantum imaging with undetected photons}

\begin{figure}
\includegraphics[width=0.6\textwidth]{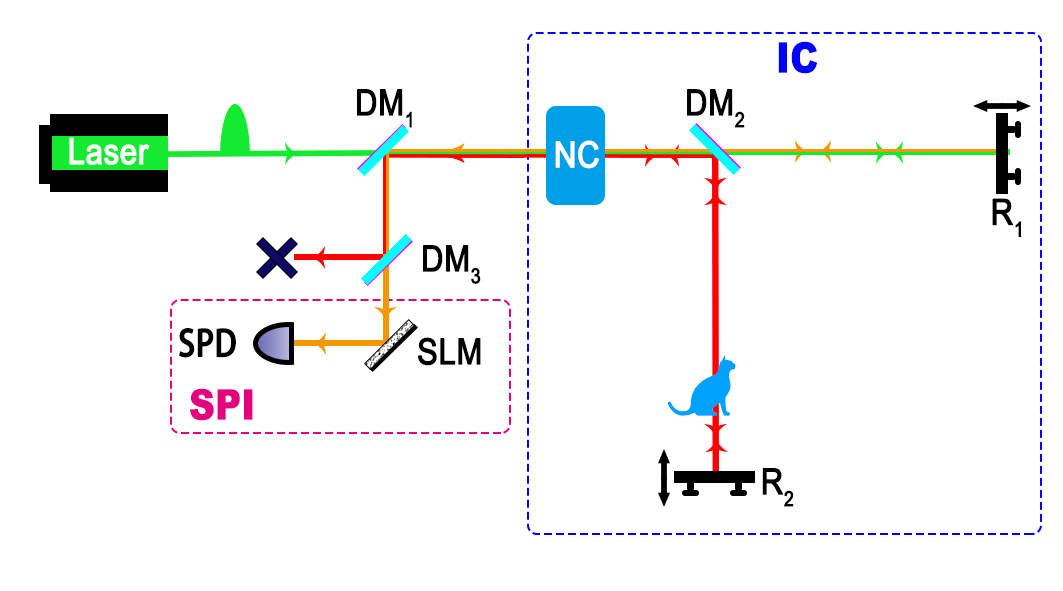}
\caption{The experimental setup of single-pixel quantum imaging with undetected photons. It consists of two modules, the induced-coherence (IC) and the single-pixel imaging (SPI) module. The idler photon is filtered out by dichroic mirror DM$_3$ and remains undetected throughout the entire imaging process. NC: nonlinear crystal; R: reflector; SLM: spatial light modulator; SPD: single-photon detector.}
\label{SPI_setup}
\end{figure}

\begin{figure}
\includegraphics[width=0.6\textwidth]{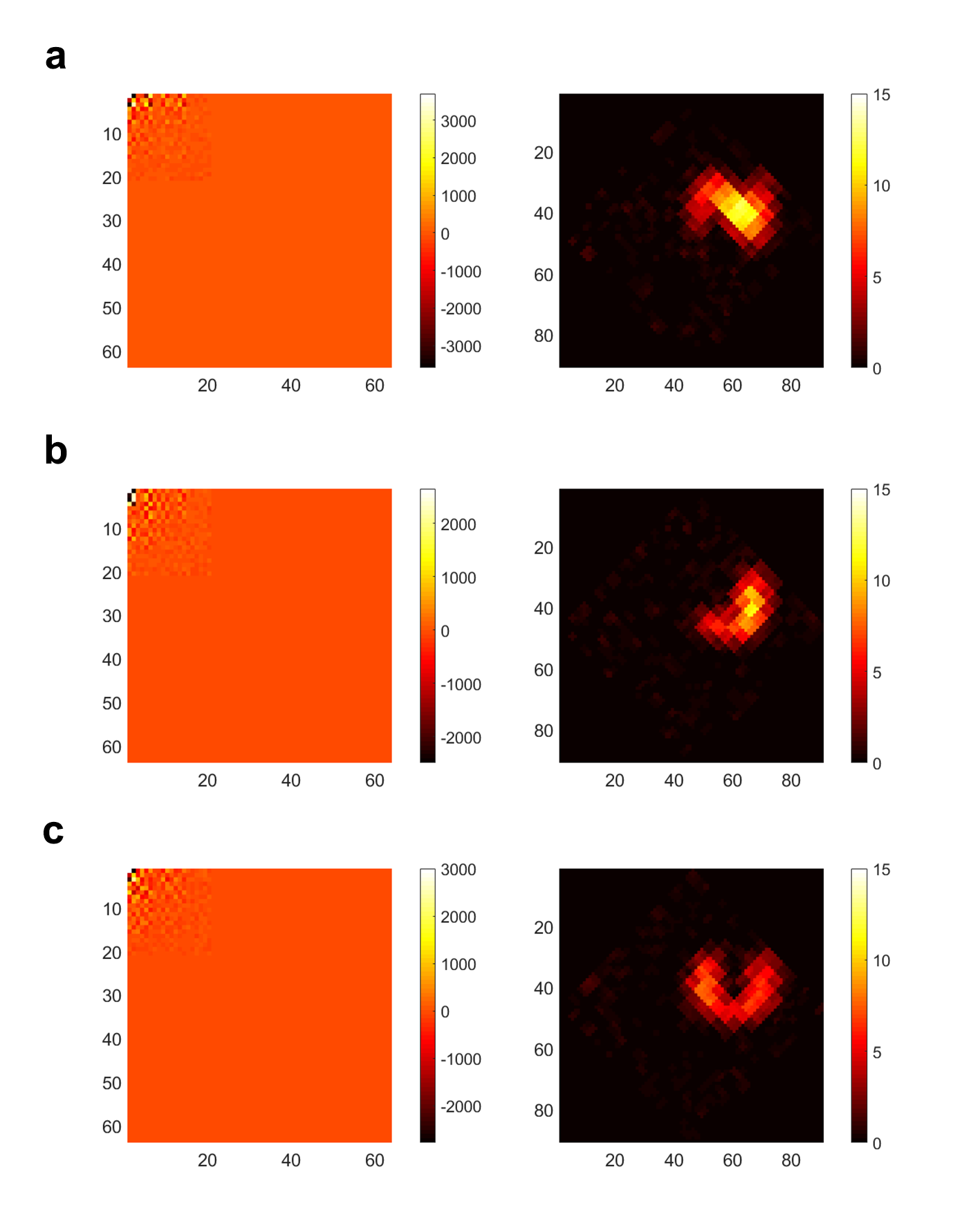}
\caption{Single-pixel quantum imaging results for the `\textbf{NJU}' plate with undetected photons. We employed 400 Hadamard masks in the single-pixel imaging process. The left and right panels are Hadamard spectra and reconstructed images of the characters `\textbf{N}', `\textbf{J}', and `\textbf{U}', respectively.}
\label{SPI_results}
\end{figure}

We also performed single-pixel quantum imaging with undetected photons. The experimental setup is shown in Fig. \ref{SPI_setup}. In contrast to the experimental system described in the main text (Fig. 1(d)), this system has no IFM module. While displaying each Hadamard mask, we recorded photon counts $C_{max}$ and $C_{min}$ at the constructive and destructive points of the signal photon interference by varying the position of mirror R$_1$. We took the difference between the maximum and minimum, $C_{diff}= C_{max}-C_{min}$, as the corresponding counts of each Hadamard mask. Each mask was multiplied by the corresponding signal-photon counts $C_{diff}$ to give a set of weighted masks. By summing all weighted masks, we reconstructed the images of the `\textbf{NJU}' plate shown in Fig. \ref{SPI_results}.

\begin{figure}
\includegraphics[width=0.5\textwidth]{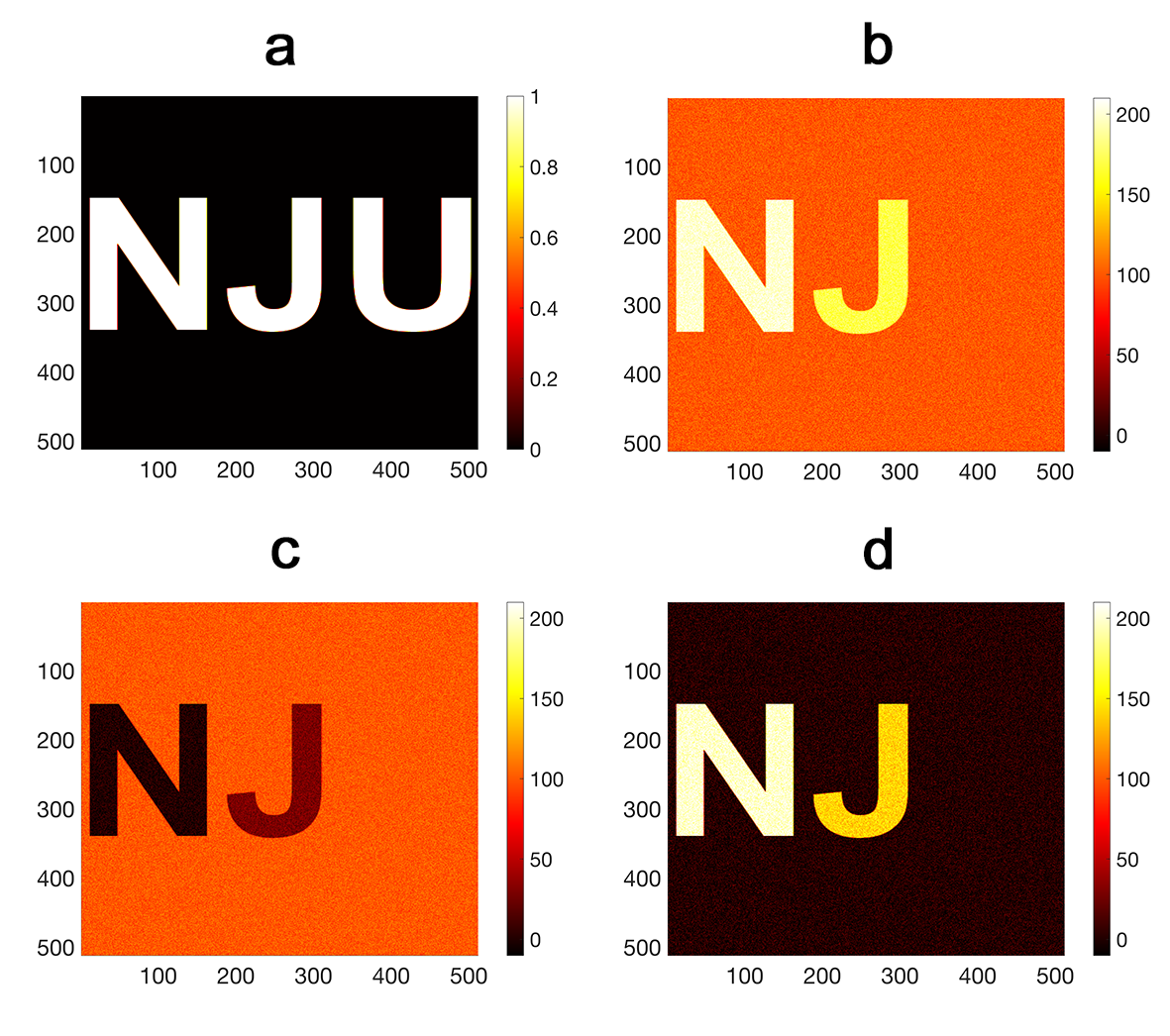}
\caption{Transparency and simulated images of the `\textbf{NJU}' logo with 512$\times$512 pixels. \textbf{a} $|T_{mn}|$ for the `\textbf{NJU}' logo transparency with `\textbf{N}' having $T=1$, `\textbf{J}' having $T=\exp(i\frac{\pi}{8})$, `\textbf{U}' having $T=\exp(i\frac{\pi}{4})$, and $T_{mn}=0$ for all other pixels; \textbf{b} and \textbf{c} are images at constructive and destructive phase points $\theta=0$ and $\theta=\pi$; \textbf{d} is the difference of (\textbf{b}) and (\textbf{c}). The difference image owns large positive values for the character `\textbf{N}' , small positive values for the character `\textbf{J}' , and near-zero values for the character `\textbf{U}' . We assume the average photon counts in each pixel is $\ave{N} = 50$.}
\label{NJU100}
\end{figure}

\begin{figure}
\includegraphics[width=0.5\textwidth]{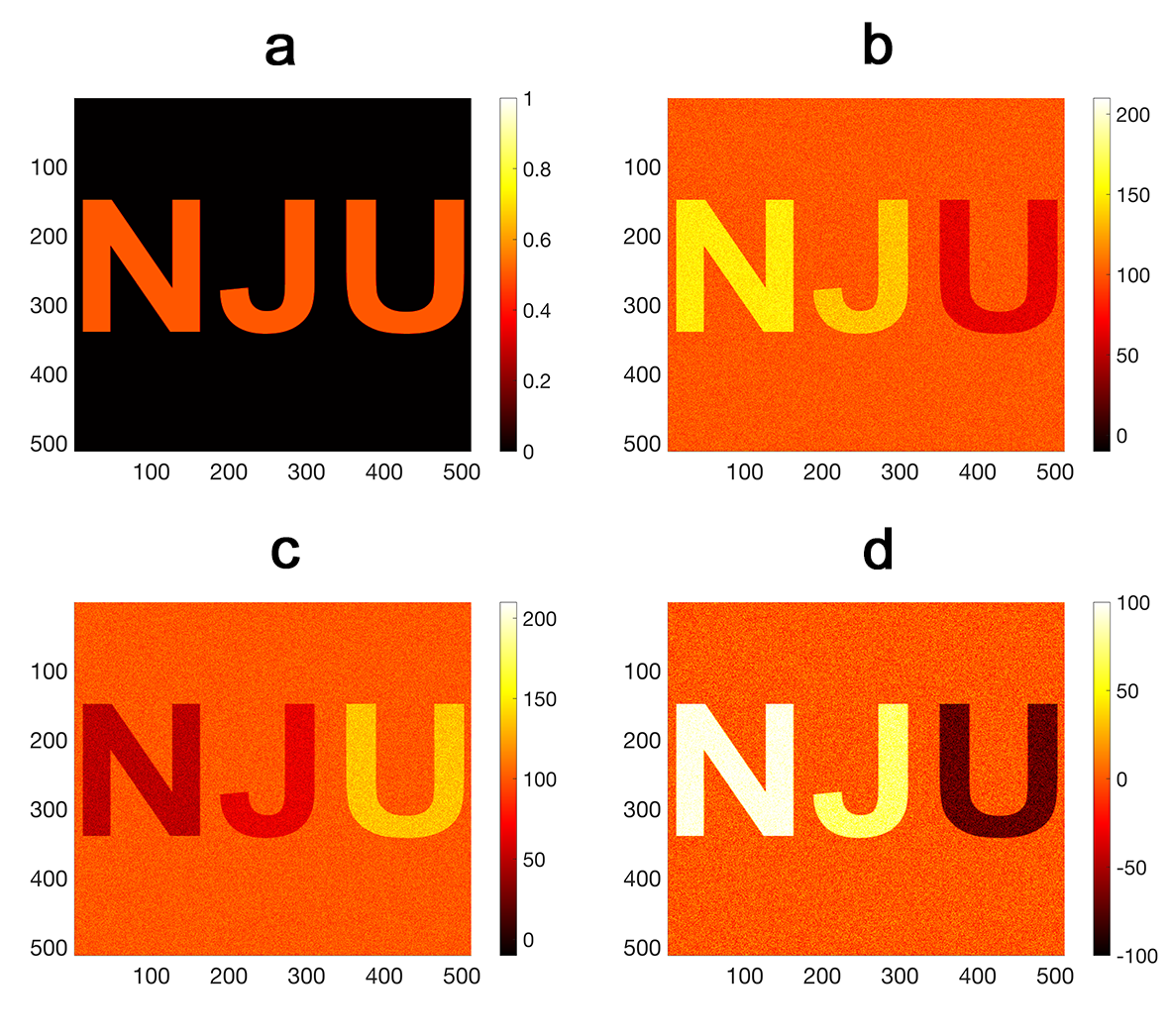}
\caption{Transparency and simulated images of the `\textbf{NJU}' logo with 512$\times$512 pixels. \textbf{a} $|T_{mn}|$ for the `\textbf{NJU}' logo transparency with `\textbf{N}' having $T=\frac{1}{\sqrt{2}}$, `\textbf{J}' having $T=\frac{1}{\sqrt{2}} \exp(i\frac{\pi}{8})$, `\textbf{U}' having $T=\frac{1}{\sqrt{2}} \exp(i\frac{\pi}{4})$, and $T_{mn}=0$ for all other pixels; \textbf{b} and \textbf{c} are images at constructive and destructive phase points $\theta=0$ and $\theta=\pi$; \textbf{d} is the difference of (\textbf{b}) and (\textbf{c}). The difference image owns large positive values for the character `\textbf{N}' , small positive values for the character `\textbf{J}' , and negative values for the character `\textbf{U}' . We assume the average photon counts in each pixel is $\ave{N} = 50$.}
\label{NJU50}
\end{figure}

Furthermore, our experimental setup can realize phase measurement due to its interference property\cite{Shapiro}. Suppose we insert one object with pixelated transmission coefficient $T=T_{mn}e^{i\delta_{mn}}$ into the idler path, then the signal interference pattern manifests as
\begin{equation}
I_{mn} = 2 \ave{N} \big[  1+T_{mn}^2 \cos(\theta +2 \delta_{mn}) \big],
\end{equation}
where $\theta$ is the relative phase of the nonlinear interferometer, $\delta_{mn}$ is the phase encoded onto the pixelated object. $\ave{N}$ denotes the average photon counts in each pixel generated by the SPDC process when induced coherence is inhibited. To demonstrate the phase imaging properties of our experimental setup, we have numerically simulated the constructive, destructive and difference images for the `\textbf{NJU}' logo with two different transmission coefficient settings, as shown in Fig. \ref{NJU100} and \ref{NJU50}. From the simulation results, we have learnt: (1) phase information is achievable within our experimental facility; (2) optical loss on the object decreases imaging visibility; (3) the difference image owns higher contrast than constructive and destructive images.

\subsection*{Supplementary Note 7. Interaction-free, single-pixel quantum imaging with undetected photons}

\begin{figure}
\includegraphics[width=0.8\textwidth]{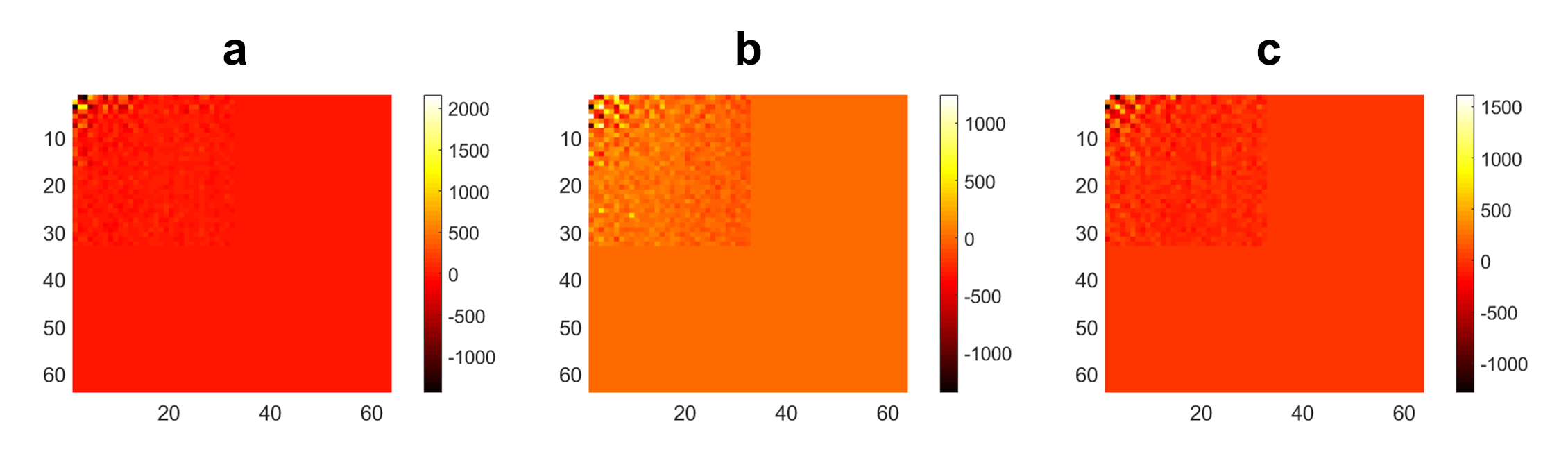}
\caption{Interaction-free, single-pixel quantum imaging results with undetected photons. \textbf{a}, \textbf{b}, and \textbf{c} are Hadamard spectra of the characters `\textbf{N}', `\textbf{J}', and `\textbf{U}', respectively. We employed 1024 Hadamard masks in the single-pixel imaging process.}
\label{Hadamard_spectra}
\end{figure}

For the interaction-free, single-pixel quantum imaging with undetected photons, we obtained the Hadamard spectra of the `\textbf{NJU}' plate shown in Fig. \ref{Hadamard_spectra}. The signal-photon counts for each Hadamard mask were filled in the corresponding position of the Hadamard spectrum. Based on measured Hadamard spectra of characters `\textbf{N}', `\textbf{J}', and `\textbf{U}', shown in Fig. \ref{Hadamard_spectra}, we constructed weighed masks by multiplying each Hadamard mask with the corresponding photon counts displayed in Hadamard spectra. Then, by summing all weighed masks, we obtained the single-pixel imaging results shown in Fig. 4 in the main text.

\end{document}